\RequirePackage{ifpdf}
\documentclass{PoS}

\graphicspath{{figs/}} 
\usepackage{mystyle} 

\usepackage{color}
\usepackage{caption, subcaption}
\let\normalcolor\relax

\newcommand{\GeV}{\text{ GeV}}
\newcommand{\subwidth}{.41\textwidth}

\title{Kaon Mixing Beyond the Standard Model}

\ShortTitle{}

\author{\speaker{A.\ T.\ Lytle}\\ Dept.\ of Theoretical Physics, 
	Tata Institute of Fundamental Research, 
 	Mumbai 400005, India \\ 
	E-mail: \email{atlytle@theory.tifr.res.in}
 	}
        
 \author{P.\ A.\ Boyle\\
 	SUPA, School of Physics, The University of Edinburgh, Edinburgh EH9 3JZ, United Kingdom\\
	E-mail: \email{paboyle@ph.ed.ac.uk} 
	}
	
\author{N.\ Garron\\ 
        School of Mathematics, Trinity College, Dublin 2, Ireland\\
        E-mail: \email{ngarron@maths.tcd.ie}   \hfill \it TCD-MATH-13-13
        }
        
 \author{R.\ J.\ Hudspith\\
 	SUPA, School of Physics, The University of Edinburgh, Edinburgh EH9 3JZ, United Kingdom\\
	E-mail: \email{R.Hudspith@sms.ed.ac.uk}
	}
 
 \author{C.\ T.\ Sachrajda \\
 	School of Physics and Astronomy, University of Southampton, 
	Southampton SO17 1BJ, United Kingdom \\
	E-mail: \email{cts@soton.ac.uk}
	}
	
\author{RBC-UKQCD Collaboration}
\abstract{
We report on an ongoing calculation of hadronic matrix elements needed to parameterize $K-\overline{K}$ mixing in generic BSM scenarios, using domain wall fermions (DWF) at two lattice spacings. 
Recent work by the SWME collaboration shows a significant disagreement with our previous results for two of these quantities. Since the origin of this disagreement is unknown, it is important to reduce the various uncertainties.
In this work, we are using $N_f=2+1$ DWF with Iwasaki gauge action at inverse lattice spacings of 2.31 and 1.75 GeV, with multiple unitary pions on each ensemble, the lightest being 290 and 330 MeV on the finer and coarser of the two ensembles respectively. 
This extends previous work by the addition of a second lattice spacing ($a^{-1}\approx 1.75 \text{ GeV}$). 
Renormalization is carried out non-perturbatively in the RI/MOM scheme and converted perturbatively to $\overline{\text{MS}}$.
}

\FullConference{The 31st International Symposium on Lattice Field Theory\\
                 July 29 -- August 03,  2013\\
                 Mainz, Germany}

\begin{document}
\section{Introduction} \label{sec:intro}
Although CP violation in kaon decays was discovered almost fifty years ago~\cite{Christenson:1964fg}, a complete theoretical computation is still missing.
As part of their kaon physics program, the RBC-UKQCD collaborations
are investigating direct and indirect CP violation~\cite{Blum:2012uk,Blum:2011ng,Boyle:2012ys} through kaon decays and neutral kaon oscillation. 
See~\cite{Janowski,Zhang,Yu,Frison,Kelly} for progress reported in this conference. 
The lattice determination of four-quark weak matrix elements, combined with experimental values of direct and indirect CP violation, provides crucial tests of the Standard Model and constrains new physics scenarios.

Kaon mixing in the Standard Model, from the effective theory viewpoint, is predominantly mediated by the single left-left $\Delta S = 2$ four-quark operator 
$\O_1 = [\bar{s} \gamma_{\mu L} d] [\bar{s} \gamma_{\mu L} d]$.
Matrix elements of this operator with kaon external states have been extensively studied in modern
lattice QCD simulations, achieving few-percent (or better) accuracy~\cite{Arthur:2012opa,Durr:2011ap,Bae:2011ff}.
Models of new physics beyond the Standard Model may be represented in the $\Delta S = 2$ sector in terms of $\O_1$ plus seven additional operators,
\begin{equation}
H_{\text{BSM}}^{\Delta S = 2} = \sum_{i=1}^{5}C_{\text{BSM}}^i(\mu) \, \O^{\Delta S = 2}_i(\mu)
+ \sum_{i=1}^{3} \tilde{C}_{\text{BSM}}^i(\mu) \, \tilde{\O}^{\Delta S = 2}_i(\mu) \,,
\end{equation}
with
\begin{align}
\O_1 &= \bigl[\bar{s}_\a \g_{\mu} (1-\g_5) d_\a  \bigr]  \bigl[\bar{s}_\b \g^{\mu} (1-\g_5) d_\b\bigr] 
\label{O1}\\
\O_2 &= \bigl[\bar{s}_\a  (1-\g_5) d_\a  \bigr] \bigl[ \bar{s}_\b   (1-\g_5) d_\b\bigr]  \\
\O_3 &=  \bigl[\bar{s}_\a (1-\g_5) d_\b  \bigr] \bigl[ \bar{s}_\b  (1-\g_5) d_\a\bigr] \\
\O_4 &=  \bigl[\bar{s}_\a  (1-\g_5) d_\a  \bigr] \bigl[ \bar{s}_\b   (1+\g_5) d_\b\bigr] \\
\O_5 &=  \bigl[\bar{s}_\a  (1-\g_5) d_\b    \bigr] \bigl[\bar{s}_\b   (1+\g_5) d_\a\bigr] \,,
\label{O5}
\end{align}
and $\tilde{\O}_{1,2,3}$ are obtained from $\O_{1,2,3}$ by $(1-\g_5) \rightarrow (1+\g_5)$.
The Wilson coefficients $C_{\text{BSM}}^i(\mu)$ are model dependent, parameterizing
the effects of heavy degrees of freedom for scales $\mu \ll M_{\text{BSM}}$.
It is thus of considerable interest to determine these matrix elements nonperturbatively in order to constrain new physics scenarios (see e.g.~\cite{Mescia:2012fg,Buras:2013ooa}).
Some recent work by other groups can be found in 
\cite{Bae:2013mja,Bae:2013tca,Carrasco:2013jaa,Bertone:2012cu}.

Results for the matrix elements of~\eqref{O1}-\eqref{O5} were reported in~\cite{Boyle:2012qb} 
and~\cite{Garron:2012ex} using $N_f=2+1$ domain-wall fermions (DWF) generated by the RBC-UKQCD collaboration with a single lattice spacing ($a^{-1} \approx 2.31 \GeV$).  This calculation has now been repeated at a second lattice spacing ($a^{-1} \approx 1.75 \GeV$).
In Sec.~\ref{sec:LEME} we present new results for the low-energy matrix elements on these configurations, which we here express in terms of ratios \cite{Babich:2006bh}
\begin{equation} \label{R_i}
R_i \equiv \frac{f_K^2}{m_K^2}\[ \frac{m_P^2 \bra{\overline{P}} \O_i \ket{P}}{f_P^2 \bra{\overline{P}} \O_1 \ket{P}} \] \,.
\end{equation}
In Sec.~\ref{sec:ren} we present some details of the renormalization procedure, in particular 
dealing with the infrared poles that arise from the RI/MOM scheme kinematics.  Sec.~\ref{sec:extrap} discusses the heavy and light mass dependences, and presents
preliminary results for our continuum-chiral global fits,
which we are working to finalize and will be presented in a future publication.

\section{Low-energy matrix elements} \label{sec:LEME}
We use the RBC-UKQCD collaboration's $N_f = 2+1$ DWF ensembles with the Iwasaki gauge
action at two lattice spacings.
These ensembles are described in detail in~\cite{Aoki:2010pe,Aoki:2010dy}, 
and the simulation parameters used in this work are summarized in Table~\ref{tab:params}.\footnote{We also refer to the ensembles  as $24^3$ ($32^3$) for $a^{-1} \approx 1.75 \GeV$ ($2.31 \GeV$).}
We use unquenched light quarks ($m_l^{\text{sea}}=m_l^{\text{val}}$) at several masses to allow extrapolation to the physical pion mass, while in the strange sector we use a partially quenched setup with single sea-quark mass
near the physical mass, and several valence masses $m_h^{\text{val}}$ with which to interpolate to the physical kaon mass.

\begin{table}[t]
\centering
\begin{tabular}{c | c | c | c | c | c | c}
extent & $a^{-1}$ [GeV] & $am^{\text{sea}}_{l} \, (= am^{\text{val}}_{l})$ & $m_\pi$ [MeV] 
 & $am^{\text{sea}}_{h}$ & $am^{\text{val}}_{h}$ & $am^{\text{phys}}_{s}$\\
\hline
 $24^3 \! \times \! 64 \! \times \! 16$ & 1.747(31) & 0.005, 0.01, 0.02 &330, 420, 560 
 & 0.04  &0.03, 0.035, 0.04 & 0.0348(11)\\
 $32^3 \! \times \! 64 \! \times \! 16$ & 2.310(37) & 0.004, 0.006, 0.008 &290, 340, 390 
 &0.03 & 0.025, 0.03 & 0.0273(7)
\end{tabular}
\caption{$N_f = 2+1$ DWF ensemble parameters used in the present study.
We use unquenched light quarks and partially quenched heavy (strange) quarks,
with the heavy masses straddling the physical strange mass.}
\label{tab:params}
\end{table}

\begin{figure}
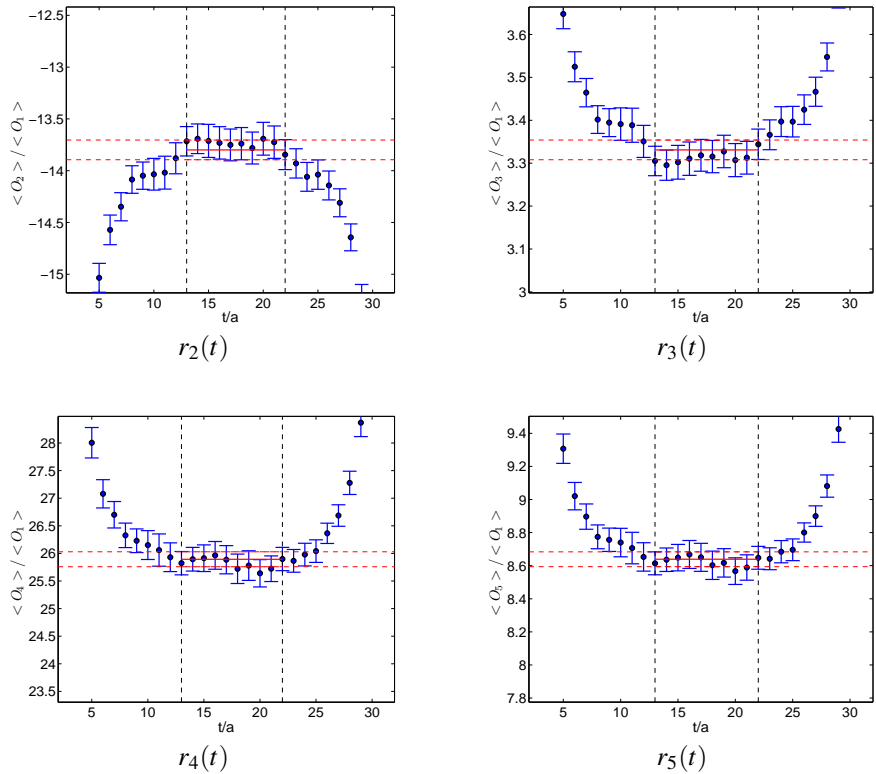

\centering
\vskip -.5cm
\begin{subfigure}[h]{\subwidth}
\includegraphics[width=\textwidth]{{{bsm_rat_with_fit_O2_susy_bas_0.005_0.0400_0.0050}}}
\vskip -.2cm
\caption*{$r_2(t)$}
\end{subfigure}
\begin{subfigure}[h]{\subwidth}
\includegraphics[width=\textwidth]{{{bsm_rat_with_fit_O3_susy_bas_0.005_0.0400_0.0050}}}
\vskip -.2cm
\caption*{$r_3(t)$}
\end{subfigure}
\vskip .3cm
\begin{subfigure}[h]{\subwidth}
\includegraphics[width=\textwidth]{{{bsm_rat_with_fit_O4_susy_bas_0.005_0.0400_0.0050}}}
\vskip -.2cm
\caption*{$r_4(t)$}
\end{subfigure}
\begin{subfigure}[h]{\subwidth}
\includegraphics[width=\textwidth]{{{bsm_rat_with_fit_O5_susy_bas_0.005_0.0400_0.0050}}}
\vskip -.2cm
\caption*{$r_5(t)$}
\end{subfigure}
\caption{Correlator ratios $r_i(t)$ showing the fits to the plateau region ($t = [13,22]$) 
on the $24^3 \! \times \! 64$ ensemble, with $am_{ud} = 0.005$
and $am_s = 0.04$.  This data was obtained using 155 configurations.}
\label{fig:r_fits}
\end{figure}

For each combination of  $m_l$ and $m_h$,
we construct correlation functions of the operators of interest between kaon interpolating operators.  We use Coulomb gauge-fixed wall sources at times $t_i$ and $t_f$ to obtain the 
correlation function (the dependence on $m_l$ and $m_h$ is left implicit)
\begin{equation}
c_i(t_i, t_f, t) =  \langle \bar{P}(t_f) \, \O_i(t) \, \bar{P}(t_i) \rangle  
\end{equation}
for all times $t$.  In the region $t_i \ll t \ll t_f$, the ratios of correlation functions
\begin{equation}
r_i(t_i, t_f, t) = \frac{c_i(t_i, t_f, t)}{c_1(t_i, t_f, t)}
\end{equation}
plateau to a constant whose value is the ratio of matrix elements
entering $R^{\text{bare}}_i(m_l, m_h)$ (cf.\ Eq.~\eqref{R_i}).
Examples of the plateau regions obtained on the $24^3$ ensemble,
for $t_i = 0$ and
$t_f = 32$, are shown in Fig.~\ref{fig:r_fits}.

\section{Renormalization} \label{sec:ren}
The bare ratios $R^{\text{bare}}_i(m_l, m_h)$ depend on the lattice regularization and
 must be matched to a continuum renormalization scheme to be useful in phenomenological applications.  
The bare and renormalized four-quark operators are related by
\begin{equation}
\O^{\MSbar}_i(\mu) = \frac{Z_{ij}}{Z_q^2}^{\MSbar} \hskip -.3cm  (\mu) \,\, \O_j^{\text{bare}} \,,
\end{equation}
where $Z_q$ is the wavefunction renormalization factor and $Z_{ij}$ is a matrix
which decomposes into a $1 \! \times \! 1$
block and two $2 \! \times \! 2$ blocks according to the $SU(3)_L  \! \times \! SU(3)_R$ chiral transformation properties of the operators $\O_{1-5}$:
$\O_1$ transforms as a (27,1), $\O_2$, $\O_3$ transform as $(6,\bar{6})$, and $\O_4$, $\O_5$ as $(8,8)$.

The matrix $Z_{ij}^{\MSbar}(\mu)$ is computed in two steps.  We first match to an intermediate
scheme, the RI/MOM scheme, which is defined both on the lattice and in the continuum.
This step is performed non-perturbatively~\cite{Martinelli:1994ty} and requires that the renormalization conditions be imposed at a scale 
\begin{equation}
\Lambda_{\text{QCD}} \ll \mu \ll \pi/a. 
\label{npr_window}
\end{equation}
In practice, the presence of pion poles makes it difficult to satisfy the condition~\eqref{npr_window} in the RI/MOM scheme and we must explicitly subtract these contributions from our data.

\begin{figure}
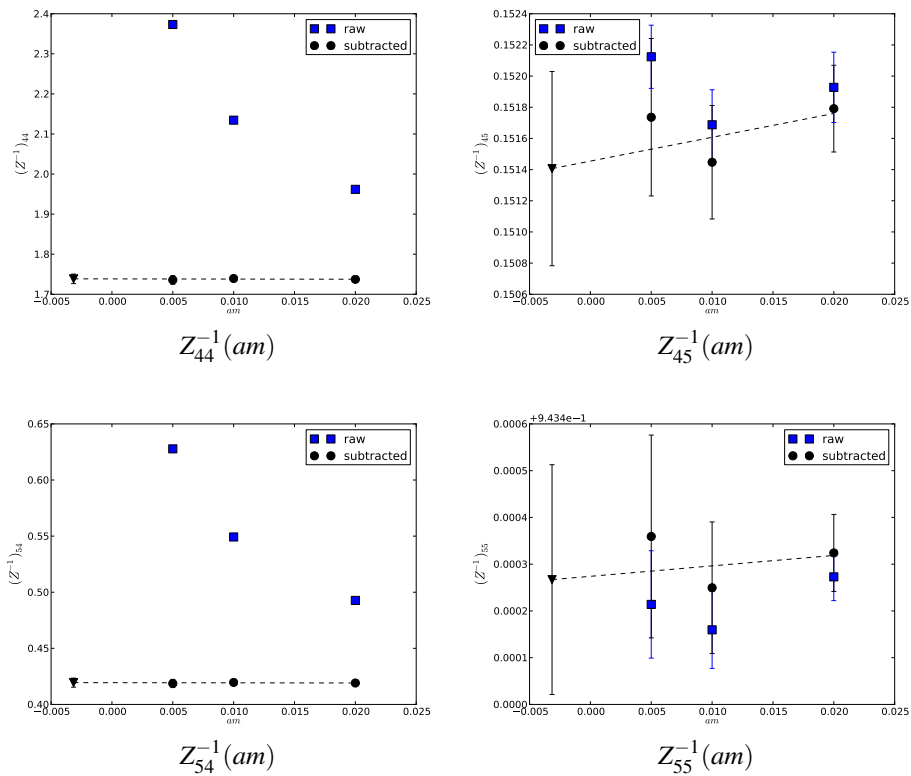

\centering
\vskip -.5cm
\begin{subfigure}[h]{\subwidth}
\includegraphics[width=\textwidth]{{{Zinv_sub_SUSY_44_exceptional_24cube}}}
\vskip -.2cm
\caption*{$Z^{-1}_{44}(am)$}
\end{subfigure}
\begin{subfigure}[h]{\subwidth}
\includegraphics[width=\textwidth]{{{Zinv_sub_SUSY_45_exceptional_24cube}}}
\vskip -.2cm
\caption*{$Z^{-1}_{45}(am)$}
\end{subfigure}
\vskip .3cm
\begin{subfigure}[h]{\subwidth}
\includegraphics[width=\textwidth]{{{Zinv_sub_SUSY_54_exceptional_24cube}}}
\vskip -.2cm
\caption*{$Z^{-1}_{54}(am)$}
\end{subfigure}
\begin{subfigure}[h]{\subwidth}
\includegraphics[width=\textwidth]{{{Zinv_sub_SUSY_55_exceptional_24cube}}}
\vskip -.2cm
\caption*{$Z^{-1}_{55}(am)$}
\end{subfigure}
\caption{NPR matrix elements of the (8,8) operators, in the RI/MOM scheme.
Blue squares are the raw data while black points show the data after pole subtraction
and extrapolation to the chiral limit.
}
\label{fig:Zsub}
\end{figure}

The renormalization factors have the form
\begin{equation}
Z^{-1}_{ij} = A_{ij} + \frac{B_{ij}}{(am)} + \frac{C_{ij}}{(am)^2} + D_{ij}(am) + \O \bigl((am)^2\bigr) \,.
\label{Z_fit}
\end{equation}
and the infrared sensitive terms $B_{ij}$, $C_{ij}$ must be subtracted.
Empirically we find that the double-pole term is benign. 
Thus  we fit $(am) Z^{-1}_{ij} \sim (am)A_{ij}  + B_{ij}$ to determine the single-pole coefficient $B_{ij}$, and then subtract this term
from~\eqref{Z_fit} to obtain subtracted elements $Z^{-1,\text{sub}}_{ij}$.
We then expect $Z^{-1,\text{sub}}_{ij}$ to behave linearly in $am$ and we use a linear
fit to determine the chiral limit.  Fig.~\ref{fig:Zsub} shows results in the (8,8) sub-matrix before and after subtraction, along with the chiral extrapolation of the subtracted data.
For those elements in which a pole is present the subtraction
procedure has a substantial effect and the resultant data is linear in $(am)$.

We also carry out  non-perturbative matching using the RI/SMOM scheme~\cite{Aoki:2010pe,Sturm:2009kb}.
This scheme has reduced sensitivity to the infrared and in particular the pole behavior
in~\eqref{Z_fit} is absent.  
It would be advantageous to use this scheme throughout the analysis, avoiding the substantial numerical subtractions e.g. seen in Fig.~\ref{fig:Zsub}.
However
the perturbative RI/SMOM $\rightarrow \MSbar$ matching factors are not presently known
for the $(6,\bar{6})$ operators.

Once the relevant renormalization factors have been non-perturbatively determined,
we convert them to the $\MSbar$ scheme using  continuum perturbation theory~\cite{Aoki:2010pe,Ciuchini:1997bw,Buras:2000if}.
Both the renormalization and matching are carried out at a scale of 3 GeV.

\section{Chiral and continuum extrapolations} \label{sec:extrap}
\begin{figure}
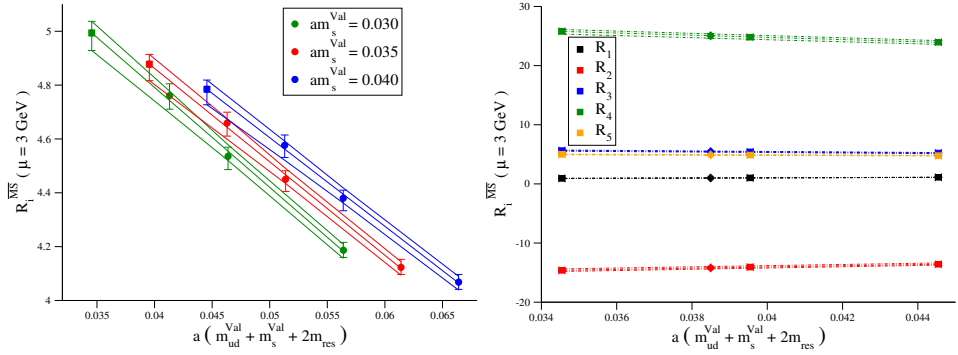

\centering
\vskip -.5cm
\begin{subfigure}[h]{\subwidth}
\includegraphics[width=\textwidth]{{{R5_24light_extrap}}}
\vskip -.2cm
\caption*{}
\end{subfigure}
\begin{subfigure}[h]{\subwidth}
\includegraphics[width=\textwidth]{{{24_extrap_noXerr_morecolor}}}
\vskip -.2cm
\caption*{}
\end{subfigure}
\caption{Quark mass dependence of the $R_i$'s.
The left panel shows the dependence of $R_5$ on the light-quark mass, for the three
values of valence strange used.  The leftmost point shows the result of linear extrapolation
to the physical light-quark mass.
The right panel shows the strange-quark mass dependence for $R_{1-5}$ (squares)
along with the interpolation point of the physical strange mass (diamond).
}
\label{fig:mass_dep}
\end{figure}

The mass dependence of $R_i(a, m_l, m_h)$ obtained from the 
light and heavy masses in Tab.~\ref{tab:params}
is used to estimate $R_i(a, m_{ud}, m_s)$ for the physical input masses 
$m_{ud}$ and $m_s$.
We find both the  $m_l$ and $m_h$ dependence of the $R_i$ to be rather mild, and
in fact consistent with linear behavior over the ranges studied. 
We linearly interpolate $R_i$ in the heavy valence mass $m_h$ to match the physical input mass $m_s$.
An example of this is shown in Fig.~\ref{fig:mass_dep} (right). 
In the light quark sector we are restricted to unphysically heavy pions and an extrapolation to lighter masses is required.  We are exploring different fit ans\"atze
and at present are using the analytic fit form. 
Results from extrapolation using this ansatz are shown in Fig.~\ref{fig:mass_dep} (left).

Having tuned the ratios to lie on the same physical scaling trajectory
for both lattice spacings, we observe discretization effects that are somewhat larger
than naively expected based on experience with $B_K$. 
To remove the leading discretization errors we employ a fit ansatz linear in $a^2$.

\begin{figure}[h]
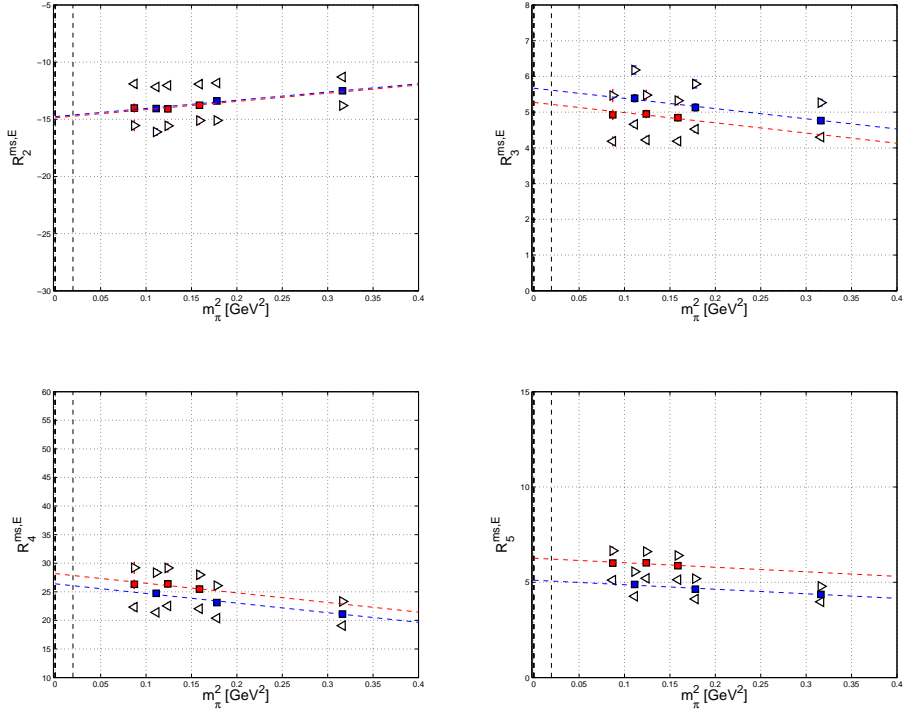

\centering
\vskip -.5cm
\begin{subfigure}[h]{\subwidth}
\includegraphics[width=\textwidth]{{{R_2_ms_E_susy_bas_no_gf}}}
\vskip -.5cm
\caption*{}
\end{subfigure}
\begin{subfigure}[h]{\subwidth}
\includegraphics[width=\textwidth]{{{R_3_ms_E_susy_bas_no_gf}}}
\vskip -.5cm
\caption*{}
\end{subfigure}
\vskip .3cm
\begin{subfigure}[h]{\subwidth}
\includegraphics[width=\textwidth]{{{R_4_ms_E_susy_bas_no_gf}}}
\vskip -.2cm
\caption*{}
\end{subfigure}
\begin{subfigure}[h]{\subwidth}
\includegraphics[width=\textwidth]{{{R_5_ms_E_susy_bas_no_gf}}}
\vskip -.2cm
\caption*{}
\end{subfigure}
\caption{(Preliminary) $R^{\MSbar}_i$ vs. $m_\pi^2$ on the $24^3$ (blue) and $32^3$ (red) ensembles.
White triangles correspond to simulated strange masses while the blue and red points are
the interpolated physical values on the respective ensembles.  
}
\label{fig:global_fits}
\end{figure}

We are also exploring global fits of the data, 
which after interpolating to the physical strange on each ensemble have the form
\begin{equation} \label{ansatz}
R_i(a, m_P^2) = R_i(0, m_\pi^2) + A_i (m_P^2-m_\pi^2) + B_i \, a^2 \,.
\end{equation} 
Minimizing $\chi^2$ across all the data determines the parameters $A_i$, $B_i$, 
and in particular our best determination of $R_i(0, m_\pi^2)$.
Fig.~\ref{fig:global_fits} shows (preliminary) results of these global fits. 
We find that the results from this method are consistent with those obtained by interpolating
for each parameter ($m_h$, $m_l$, $a^2$) in succession. 

\section{Conclusions} \label{sec:concl}
We  have computed kaon matrix elements of BSM $\Delta S = 2$ operators using $N_f = 2+1$ DWF at two lattice spacings, and at a variety of (unquenched) light quark masses and (partially quenched) strange quark masses. 
Renormalization is carried out non-perturbatively in the RI/MOM scheme and then converted
to $\MSbar$ via perturbation theory.
This renormalization scheme required a large
pion pole subtraction that is a source of systematic error.
We are currently finalizing the global continuum-chiral fits to our results  and will present the results in a forthcoming publication.

New RBC-UKQCD ensembles have been generated at the same lattice spacings,  but with spatial extents
of $48^3$ and $64^3$ and masses at the physical point~\cite{Mawhinney}.
We are running measurements on these and plan to add this data to remove uncertainty from the extrapolation ansatz~\eqref{ansatz}.
We have also carried out renormalization in the RI/SMOM scheme
which will remove the systematic error due to pole subtractions.

\section{Acknowledgements} 
We warmly thank the conference organizers,
and our colleagues in the RBC and UKQCD collaborations for their helpful input and sharing of resources.
C.T.S. acknowledges STFC Grant ST/G000557/1.  
The authors gratefully acknowledge computing time granted through the STFC funded DiRAC facility (grants ST/K005790/1, ST/K005804/1, ST/K000411/1, ST/H008845/1). P.A.B. acknowledges support from STFC Grant ST/J000329/1 and was also supported by the European Union under the Grant Agreement number 238353 (ITN STRONGnet).


\let\oldbibliography\thebibliography
\renewcommand{\thebibliography}[1]{%
  \oldbibliography{#1}%
  \setlength{\itemsep}{-1pt}%
}

\end{document}